\documentclass[11pt,twoside]{article}

\usepackage{asp2006}
\usepackage{epsf}
\usepackage{psfig}
\usepackage{lscape}

\begin{document}
\title{On classical and pseudobulges: The nature of bulges of early-type spirals}   
\author{Reynier F. Peletier}   
\affil{Kapteyn Astronomical Institute, University of Groningen, Postbus 800, 9700 AV Groningen, The Netherlands}    

\begin{abstract} 
Historically, galactic bulges are thought to be elliptical galaxy-like objects sitting in the middle of a generally larger disk. There are, however, more and more claims that some bulges are much more similar to disks. John Kormendy has named these bulges pseudo-bulges. In this paper I discuss some recent integral field spectroscopy of the SAURON collaboration on a sample of 24 Sa and Sab galaxies. Using their 2-dimensional maps of the stellar velocity, velocity dispersion, and absorption line strength, it is now much easier to understand the nature of nearby galactic bulges. I review several aspects of bulges, namely the surface photometry, stellar kinematics, stellar populations, presence of interstellar matter, and their behavior in the fundamental plane of early-type galaxies.
\end{abstract}


\section{Introduction and definition of a bulge}
One of the remarkable features of an image of a spiral galaxy is its central light concentration. The recent Spitzer Space Telescope images once again showed some beautiful examples of this, in particular the Andromeda galaxy (SSC 2006-14), the Sombrero (SSC 2005-11), M 51 (SSC 2004-19) and NGC 7331 (2004-12). On the images one sees that every galaxy has a central light concentration due to stellar light with relatively few features due to dust and star formation. This light concentration is known as a galactic bulge. In the old days bulges were known as old, metal rich, galaxy components (Whitford 1978). Nowadays, there are strong indications that there is at least a large population of galactic bulges which also contain younger stars. Details about this can be found in the long review by Kormendy \& Kennicutt (2004, hereafter \textbf{KK}). Since research on bulges is making fast progress, however, I would like to revisit the subject in the light of new observations, especially from the SAURON instrument (Bacon et al. 2001), and discuss what we know about bulges, and what we can deduce about their formation and evolution. 

Observations from the SDSS Survey (Baldry et al. 2004) show that there are two
main groups of galaxies: those on the red sequence, and a group of bluer
galaxies, the so-called blue cloud. The red sequence contains the elliptical galaxies in the nearby Universe. Mainly because of their similarities with elliptical galaxies, galactic bulges are traditionally thought to be elliptical galaxies in the middle of spiral disks (\textbf{KK}), and therefore thought to be the product of early dissipative collapse (Eggen, Lynden-Bell \& Sandage 1962) and mergers (Toomre 1977), around which later a disk formed. In more recent years, however, alternative theories have been developed, in which bulges formed from disk material (Pfenniger \& Norman 1990 and others). Evidence that this process of so-called secular evolution is really happening is summarized in \textbf{KK}. 

To have a clean discussion it is important to first define what we mean when we
talk about a galactic bulge. Is this the same as what we measure? Traditionally,
people define the bulge as the light excess above an exponential disk (Freeman
1970). This is a purely observational definition, applicable to all galaxies. We
call this the \textit{photometric} decomposition. It suffers, however, from the
fact that some disks cannot be fitted well by exponential surface brightness
distributions, e.g. in the presence of bars, and that we don't know what this bulge means physically. Kent (1986), therefore introduced the \textit{morphological} decomposition, assuming that bulge and disk have different flattening. His method separates non-parametrically two components with different projected ellipticities. This method also has its problems: galaxies, especially their inner regions, have axis ratios that vary with radius. Also, there are spirals for which the axis ratio in the central regions is the same as in the outer disk (\textbf{KK}). For face-on galaxies this method is in any case unusable. 
If a spiral galaxy would consist of a slowly rotating elliptical galaxy surrounded by a fast rotating exponential disk one could also think of doing the decomposition using the kinematics. Due to the lack of high quality kinematic profiles, this \textit{kinematic} decomposition has not been tried often in the literature, but it has been mentioned (e.g. Kormendy 1993). I can conclude that nowadays photometric decomposition is by far the most common. It might actually be preferable to use the kinematic decomposition, since using this method it would be much easier to understand the origin of bulges.

So, are bulges indeed similar to ellipticals, as I myself have claimed a number of years ago (Peletier et al. 1999)? On first view one might think that it is hard to include galaxies such as NGC 4736 (Kormendy 1993), which have spiral structure in the bulge region, and which rotate faster than one would expect from an elliptical, in this category. Also, bulges of late-type spirals do not follow the de Vaucouleurs r$^{1/4}$ law (Andredakis et al.) and have bright nuclear star clusters (e.g. B\"oker et al. 2002). What fraction of the bulges of early-type spirals (type S0a-Sb) looks like an elliptical? In the rest of this paper I will discuss a few aspects of the observations of bulges that maybe can bring us somewhat closer to answering this question and understanding the nature of bulges.

\section{Surface brightness profiles}

Surface brightness profiles of bulges are well fitted by r$^{1/n}$ S\'ersic
distributions (for a historic review see \textbf{KK}). The S\'ersic index $n$
correlates well with morphological type (Andredakis, Peletier \& Balcells 1995) and bulge to
disk ratio, giving values of around 4 and higher for ellipticals and S0's,
decreasing for values around 1-2 for Sc's. Balcells et al. (2003, 2007) noted
that surface brightness profiles of early type spirals could be fitted better if
an additional central point source is included. For a sample of inclined spirals
with inclination larger than 50$^{\rm o}$ of type S0-Sbc (the Balcells/Peletier
sample) the S\'ersic index of the bulge rarely exceeds 3, even for S0's. Central
sources are also present very often present in late type spirals. B\"oker et al.
(2002) find that 77\% of their sample of 77 late-type spirals contains a nuclear
cluster, which is often resolved. In intermediate-luminosity Virgo cluster ellipticals Cot\'e et al. (2004) reports a detection rate of 82\%. Motivated by theory Carollo et al. (2002) classifies her sample of bulges into so-called r$^{1/4}$- or classical bulges and exponential bulges. In agreement with the scatter found by Andredakis et al. (1995) both classical and exponential bulges can be found at most morphological types, although for the latest types only exponential bulges are found. However, is it reasonable to make this separation into 2 groups, since the underlying distribution of S\'ersic indices is continuous? To justify this, other, independent arguments should also be used. 

Interesting to note about the surface photometry is that the scales of bulges
and disks are correlated (Courteau, de Jong \& Broeils 1996). This implies that the bulge is strongly influenced by its surrounding disk, or vice-versa.

\section{Boxy "bulges"}

A significant fraction of the edge-on galaxies contains a so-called \textit{boxy} or peanut-shaped bulge, i.e., the outer isophotes of the bulge appear strongly boxy (see e.g. Jarvis 1986). It is well-established that these are bars seen edge-on. N-body simulations show that they are due to bar-orbits buckling out of the plane (see e.g. Athanassoula 2005). Also, observationally, Kuijken \& Merrifield (1995) have shown that along the peanut they see velocity-splitting, characteristic for the presence of a bar (see also Chung \& Bureau 2004). Moreover, in the azimuthally averaged surface brightness profiles, the boxy bulge is generally not identified. Since these arguments show that boxy bulges are related to bars, which are an integral part of the flat disk, I prefer to conclude that their name is very misleading. Although boxy, these components are part of the disk and should maybe be called the boxy region. They have little to do with the component in exces of the exponential disk, the bulge.

\section{Stellar kinematics}

The (v/$\sigma$) -- $\epsilon$ diagram has been a very powerful tool to characterise the nature of galactic bulges. Kormendy \& Illingworth (1982) showed that bulges fall close to the oblate rotator line in this diagram, the line of an isotropic object flattened by its rotation. They, and also Davies et al. (1983), showed that kinematically, these bulges behave the same as intermediate mass ellipticals. The bulges for which kinematics was available at that time were all large bulges of early-type spirals. Later, Kormendy could add other bulges to the diagram, and found that some had larger (v/$\sigma$) than expected for an isotropic rotator, showing more disk-like kinematics.  In recent years, the whole picture is becoming much more clear. Many galaxies have been discovered in which a thin, rotating disk is dominating the light in the very inner parts, accompanied by a local minimum in the velocity dispersion. The first observed cases of central
velocity dispersion minima date back to the late 80s and early 90s (e.g., Bottema 1989, 1993). Several others were found from long-slit data (Emsellem et al. (2001), M\'arquez et al. 2003). SAURON observations showed that 13 out of 24 Sa and Sab galaxies showed a central local minimum in the velocity dispersion (F\'alcon-Barroso et al. 2006). We see a weak trend that galaxies with sigma-drops are younger than those without (Peletier et al. 2007). The sigma-drops are probably due to central disks that formed from gas falling into the central regions through a secular evolution process. Simulations show that the disks will remain in place for a long time after they have been formed (Wozniak \& Champavert 2006), although they will slowly heat up with time. Such central disks are also known among elliptical and S0 galaxies (e.g. NGC 4526, Emsellem et al. 2004, NGC 7332, Falc\'on-Barroso et al. 2004), but rare. 

\section{Stellar populations}

Up to 1990, very little was known about the stellar populations of galactic
bulges, apart from our own Galaxy, which was thought to consist of old,
super-metal rich stars (e.g. Rich 1988). Balcells \& Peletier (1994) showed, by
looking at the bulge on the side of inclined galaxies not obscured by the
dustlane, that the colours were the same or bluer as elliptical galaxies of the
same luminosity. Later, they showed, from colour-colour diagrams, that these
galaxies were old and almost all coeval, 9 $\pm$ 2 Gyr (Peletier et al. 1999).
Bulges are not visible in colour maps, implying that bulges and inner disks have
very similar stellar populations. When put on the fundamental plane of
early-type galaxies the same sample falls on the line defined by early-type
galaxies in the Coma cluster (J\o rgensen, Franx \& Kjaergaard  1996), confirming that they are old.

If one, however, looks at the Mg$_2$ - $\sigma$ relation, a different result is obtained (Peletier et al. 2007). Although several samples of bulges fall on the Mg$_2$ - $\sigma$ relation of early-type galaxies in Coma, a large fraction of the sample of early-type spirals of Prugniel, Maubon \& Simien (2001) has a lower Mg$_2$ value for the same $\sigma$. Also, several bulges of the SAURON sample of Peletier et al. (2007) fall below the relation for Coma. The difference between the samples that fall on the line, and those that fall partly below it is that the samples that fall on the line consist either of inclined galaxies, or mainly of S0 galaxies. Apparently, bulges of samples for which the inclination is unbiased can have a large range in ages. This can be checked on the SAURON line strength maps (especially the H$\beta$ absorption, Peletier et al. 2007), which confirm that the galaxies that lie far from the Mg$_2$ - $\sigma$ for Coma galaxies are indeed young. This also explains the large range in ages found by others (e.g. Proctor \& Sansom 2002, Moorthy \& Holtzman 2006). Our late-type spirals of Ganda et al. (2006), observed with SAURON, all fall below the Coma relation.

More insight in the stellar populations can be obtained when one looks at line strength gradients. Recently, Jabonka, Gorgas \& Goudfrooij (2007) (see also Gorgas, Jablonka \& Goudfrooij 2007) obtained line strength gradients of a sample of 32 edge-on galaxies, and showed that outside the central dust-lane Mg$_2$ decreases as a function of distance above the plane in almost all galaxies. The same is the case for all their other indices, except for the Balmer indices. This can be explained if metallicity is decreasing vertically outwards. For the SAURON Sa-Sab galaxies, however, we find that gradients can be both positive and negative. Looking at the line strength maps this is directy understood to be due to the fact that these galaxies contain regions of recent star formation associated to dynamical resonances, i.e. rings and spiral arms. If a galaxy has star formation in the center, the Mg$_2$ gradient is positive, while in the case of no star formation, or a star forming ring at the edge of the bulge the gradient is negative. A similar behavior is seen in Moorthy \& Holtzman (2006). Again, a difference is seen between inclined and unbiased samples. When one looks above the plane, bulges appear old. When, on the contrary, one looks in the plane itself, there are many regions with young stars.

\section{Gas, dust and spiral structure in bulges}

Spiral galaxies are dusty objects in their centers. Sb and Sc galaxies have on the average about 1 mag of extinction in $B$ in the central regions, and are basically transparent at 3 effective disk scale lengths (Giovanelli et al. 1995). Sa galaxies contain less extinction. Martini et al. (2003) show a large collection of central HST colour maps of nearby spirals. The maps show that the large majority of galaxies have extinction in the central regions. There are galaxies with nuclear rings very close to the center, and well within the bulge. There are highly inclined galaxies with smooth colour maps on the side of the galaxy away from the dust lane. Many galaxies show spiral structure in their inner regions, a sign that the dust is distributed in a plane. Regan et al. (2006) and Bendo et al. (2007) show that many spirals, also of type Sa, show the presence of 8 $\mu$m and 24 $\mu$m emission in their central regions, as well as molecular gas, as traced by CO. 
colour maps of edge-on galaxies (e.g. Howk \& Savage 2000) show that dust
extinction can be found high above the plane, but also that the dust seems to be
associated to the disk only. The fact that colour maps of bulges of galaxies
that are not obscured by the disk, i.e. at inclinations between 50 and 85$^{\rm
o}$, are smooth also strongly suggests that the dust is found in the plane only.
Since dust and gas are often seen together (see e.g. Kent, Dame \& Fazio 1991,
Sarzi et al. 2006) I expect that the ionised, neutral and molecular gas will also be found near the plane only, where it can be in equilibrium. 

I would like to give a word of caution here about determine stellar population parameters from colours in the presence of extinction. When one is reasonably sure that the colours are not or only slightly affected by dust (e.g., Peletier et al. 1999) stellar populations can in principle be derived from colours. The same paper also shows, however, that the \textit{central} colours of the Balcells \& Peletier sample are so red that they cannot be fitted by any realistic stellar population model. Correcting for extinction here is very difficult, since the error in the correction is often so large that one has no distinguishing power left in corrected colour. For dusty galaxies it is also very tricky to mask out regions with a lot of extinction and determine the colours in places where one thinks there is no extinction (e.g. Carollo et al. 2007), and most certainly the result will not be reproducible by others. SAURON line strength maps (e.g. Ganda et al. 2007) should offer a reasonable alternative.

\begin{figure}
\begin{center} 
\psfig{figure=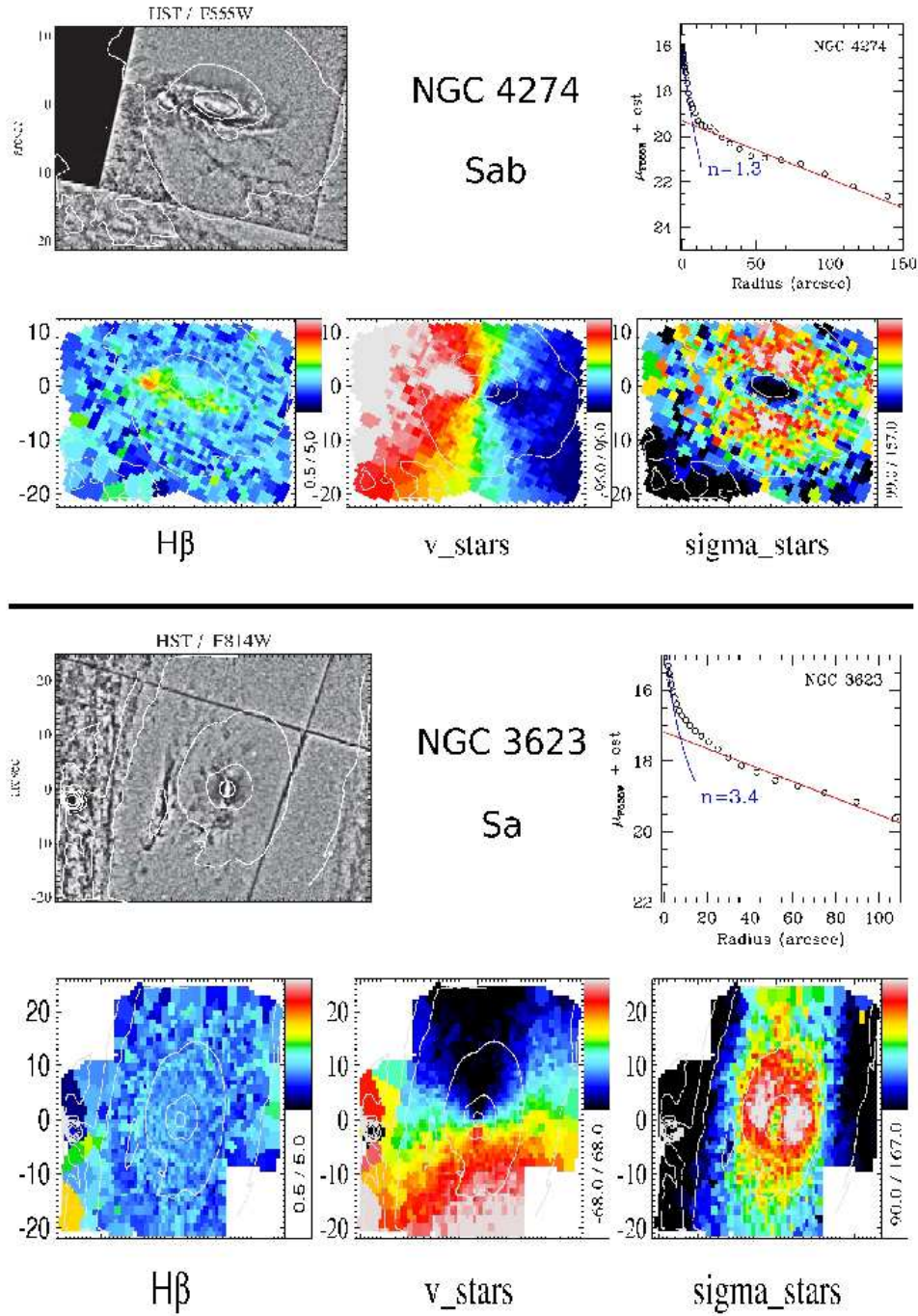,width=13.0cm}
\end{center} 
\caption[]{Diagnostic observations of NGC 4274 and NGC 3623, two early-type spirals. For each galaxy is shown in the top left:
unsharp-masked F555W HST image, showing the places of non-negligible extinction
(from Falc\'on-Barroso et al. 2006). Top right: major axis surface brightness
profile, from the same HST data. A bulge-disk decomposition is also shown, with
an exponential disk and a S\'ersic bulge. Bottom row (from left to right):
H$\beta$ absorption line map (Peletier et al. 2007),  stellar velocity and velocity dispersion map
(Falc\'on-Barroso et al. 2006). Overlayed on all maps is the
reconstructed SAURON intensity.}
\end{figure}

\section{Our new picture of galactic bulges}

From the stellar population distributions,  by comparing samples
uniformly distributed in inclination with samples biased towards  high
inclination we infer that galactic bulges have more than one physical component:
generally they contain a slowly-rotating, elliptical-like component, and one or
more fast-rotating components in the plane of the galaxy, all co-existing in the same galaxies.
This picture also
nicely explains the fact that bulge populations in general are very similar to
those in the disc (e.g. Peletier \& Balcells 1996).  
It is also strongly supported by the kinematics. In more than half of the SAURON
early-type spirals sigma-drops occur. These correspond to central disks, which sometimes 
contribute most of the light corresponding to the bulge resulting from the photometric 
decomposition. HST images shows that the central disks correspond to dusty regions, 
often showing spiral structure. They 
also often, but not always, contain younger stellar populations than the regions outside the 
central disks.

In Figure 1 we give an example two galaxies: one that is classified, based on the photometric decomposition, as a pseudo-bulge, and one of a classical bulge. The first example, NGC 4274, for which the surface brightness profile is fitted best by a S\'ersic profile with $n$=1.3, has a strong sigma-drop. In the 
region where it dominates the stars rotate fast and have a
low velocity dispersion (SAURON). Such an object would be called a pseudo-bulge by \textbf{KK}. The bulge is dusty, has spiral structure (HST image) and shows younger stellar populations (SAURON). Note also that as one goes out on the minor axis, the stellar distribution becomes smooth, and the line strengths show values corresponding to old stellar populations. This region corresponds to an elliptical-like bulge,
to which \textbf{KK} refer as a classical bulge. NGC 4274 therefore contains a pseudo-bulge \textit{AND} a classical bulge. Making the classical bulge a bit larger, and the pseudo-bulge smaller, one gets an object such as NGC 3623 (bottom of the Figure). Here the classical bulge dominates the light, apart from the very inner regions. A sigma-drop is seen, but weaker than in NGC 4274 (Falc\'on-Barroso et al. 2006). The S\'ersic fit for the bulge gives $n$=3.4. Dust is associated with the central disks, but the SAURON maps show that the stellar populations in it are not different from the classical bulge outside it. The comparison between NGC 3623 and NGC 4274 shows that both objects are very similar, but that the inner disk to elliptical bulge ratio in both galaxies is different.

The Galactic Bulge also fits well into this picture, of having a hot and a cold component living together in the same galaxy. Zoccali et al. (2003) find that stars
in the Galactic Bulge, measured 6 degrees above the Galactic plane, are as old as Galactic
globular clusters, at least 10 Gyr. This indicates that our Galaxy probably has an old
elliptical-like bulge component.  We do not know of a component
with disc-like kinematics in our Galaxy, but near the Galactic centre stars are currently
being formed at a fast rate. A nuclear stellar disk is found with a scale of about 100 pc (Launhardt et al. 2002), as well as a nuclear cluster.

Fisher (2006) reports one more
piece of evidence. In galaxies which he classifies as having a pseudo-bulge, the central
star formation rate, as obtained from the Spitzer 3.6 - 8.0 $\mu$m colour, is higher than
in galaxies with classical bulges. This shows that the Spitzer colour measures similar
things as optical absorption line indices. 
Recently, Drory \& Fisher (2007) classify the bulges in a sample of low inclination S0-Sc galaxies into 
classical and pseudo-bulges. If they see signs of disk-features, such as nuclear bars, spirals or rings the bulge is called a pseudo-bulge, otherwise a classical bulge. Apart from the caveat that many galaxies probably contain both types of bulges, their result that their bulge-classification correlates with colour
is not surprising, since these nuclear features are indicators of recent star formation. If the bulge looks featureless, it has been shown in the past that the stellar populations are old.

It is clear that our knowledge about bulges is increasing rapidly. It is now up to the observers to firmly establish these new ideas, and to the theorists to not only understand what we observe for bulges, but also to understand the implications for galaxy formation in the Universe. 

\acknowledgements 

I would like to thanks the organisers of this pleasant meeting, and wish John Beckman many more interesting years in Astronomy. I am grateful for the help of my collaborators Marc Balcells, Jes\'us Falc\'on-Barroso and Katia Ganda in particular.



\end{document}